\documentclass[]{spie}  

\usepackage{amsmath,amsfonts,amssymb}
\usepackage{graphicx}
\usepackage[colorlinks=true, allcolors=blue]{hyperref}
\usepackage{caption}
\usepackage{subcaption}
\usepackage{afterpage}
\usepackage[colorinlistoftodos]{todonotes}
\usepackage{tabularx}
\usepackage{algorithm}
\usepackage[noend]{algpseudocode}
\usepackage[colorlinks=true, allcolors=blue]{hyperref}
\usepackage[normalem]{ulem}
\usepackage{afterpage}
\usepackage{stackengine}

\newcommand{\universe}[1]{\mathbb{R}^{#1}}
\newcommand{\setA}{\mathbf{U}}
\newcommand{\setB}{\mathbf{V}}

\title{	Experimental demonstration of photonic phase correctors based on grating coupler arrays and thermo-optic shifters}

\author[a]{Momen Diab}
\author[b]{Ross Cheriton}
\author[a,c]{Jacob Taylor}
\author[a]{Dhwanil Patel}
\author[a]{Libertad Rojas}
\author[a]{Mark Barnet}
\author[a]{Polina Zavyalova}
\author[b]{Dan-Xia Xu}
\author[b]{Pavel Cheben}
\author[b]{Siegfried Janz}
\author[b]{Jens H. Schmid}
\author[a,c]{Suresh Sivanandam}
\affil[a]{Dunlap Institute for Astronomy and Astrophysics, University of Toronto, Toronto, Ontario, Canada}
\affil[b]{Quantum and Nanotechnologies Research Centre, National Research Council Canada, Ottawa, Ontario, Canada}
\affil[c]{David A. Dunlap Department of Astronomy and Astrophysics, University of Toronto, Toronto, Ontario, Canada}

\authorinfo{Further author information: (Send correspondence to M.D.)\\M.D.: E-mail: momen.diab@utoronto.ca}

\begin{document} 
\maketitle

\begin{abstract}


In ground-based astronomy, the ability to couple light into single-mode fibers (SMFs) is limited by atmospheric turbulence, which prohibits the use of many astrophotonic instruments. We propose a silicon-on-insulator photonic chip capable of coherently coupling the out-of-phase beamlets from the subapertures of a telescope pupil into an SMF. The photonic integrated circuit (PIC) consists of an array of grating couplers that are used to inject light from free space into single-mode waveguides on the chip. Metallic heaters modulate the refractive index of a coiled section of the waveguides, facilitating the co-phasing of the propagating modes. The phased beamlets can then be coherently combined to efficiently deliver the light to an output SMF. In an adaptive optics (AO) system, the phase corrector acts as a deformable mirror (DM) commanded by a controller that takes phase measurements from a wavefront sensor (WFS). We present experimental results for the PIC tested on an AO testbed and compare the performance to simulations.

\end{abstract}


\section{Introduction}
Aberrations introduced by Earth's atmosphere limit the resolution of astronomical telescopes, reducing the resolution of imagers, the resolving power of slit spectrographs, the contrast in coronagraphs, and the coupling efficiency into fibers and astrophotonic instruments \cite{jovanovic_2023_2023}.
Adaptive optics (AO) systems have increasingly been used to compensate for phase errors in optical wavefronts caused by atmospheric turbulence. In AO systems, a wavefront sensor (WFS) samples the collected wavefront and measures the distortions, while a wavefront corrector (WFC) flattens the wavefront based on commands received from a controller fed by the WFS signals. Typically, a Shack-Hartmann wavefront sensor (SH-WFS) or a pyramid sensor is used to measure the wavefront. Both of these classical types of WFSs differ in terms of dynamic range, sensitivity, and spatial resolution, but they all require a bulk optic and a detector to split and sense the incoming wavefront. They also work on the pupil plane, making them blind to certain modes and non-common path aberrations (NCPAs) introduced between where the light is sampled and where the science instrument is placed. New photonic concepts for WFSs, with most based on photonic lanterns\cite{norris_all-photonic_2020,diab_modal_2019}, have been suggested as additional sensors in the AO system to measure the blind modes and the NCPAs, since they sense the complex field at the focal plane where the petal modes and the NCPAs are detectable.  

Deformable mirrors (DMs) have always been the method of choice for the WFC in astronomy. Multiple implementations of the DM have been developed, but voice coils and micro-electromechanical systems (MEMS) are by far the most common. These two options fulfill different requirements in terms of mechanical stroke, actuator density, i.e., spatial resolution, and temporal response, but they are costly devices that cannot be efficiently multiplexed. A cost-effective photonic alternative to DMs with a multiplexing advantage could allow scaling up the number of corrected objects in fibered multi-object spectrographs (MOS) fed by multi-object AO (MOAO) systems with a high count of targeted objects, e.g., $\sim 100$s of objects. In free-space optical (FSO) communication systems, the high temporal bandwidth of photonic components, in addition to their low cost and larger stroke, is important for tracking fast-slewing low-Earth orbit (LEO) satellites, where coupling the distorted laser beam into single-mode fibers (SMFs) is essential to enable amplification and detection. \cite{thompson_nasas_2023}        

We fabricated a silicon-on-insulator (SOI) photonic integrated circuit (PIC) that corrects optical wavefronts by coupling the light focused by a microlens array (MLA) into single-mode waveguides using an array of grating couplers. 
We previously reported the theoretical framework and simulation results for the proposed concept \cite{diab_photonic_2022, patel_end--end_2024}. Here, we present an experimental demonstration of a photonic corrector with a $2\times 2$ array as a proof of concept and a pathfinder for next-generation larger arrays. In Sec. \ref{sec:pic}, the photonic components of the PIC are described, while the optical components of the rest of the setup are detailed in Sec. \ref{sec:setup}. Section \ref{sec:dms} provides a comparison between photonic phase correctors and DMs based on theoretical and simulation results.  The experimental results are given in Sec. \ref{sec:results}, followed by a discussion in Sec. \ref{sec:discussion}.      

\section{Photonic Integrated Circuit}
\label{sec:pic}

\begin{figure}
    \centering
    \includegraphics[width=0.5\linewidth]{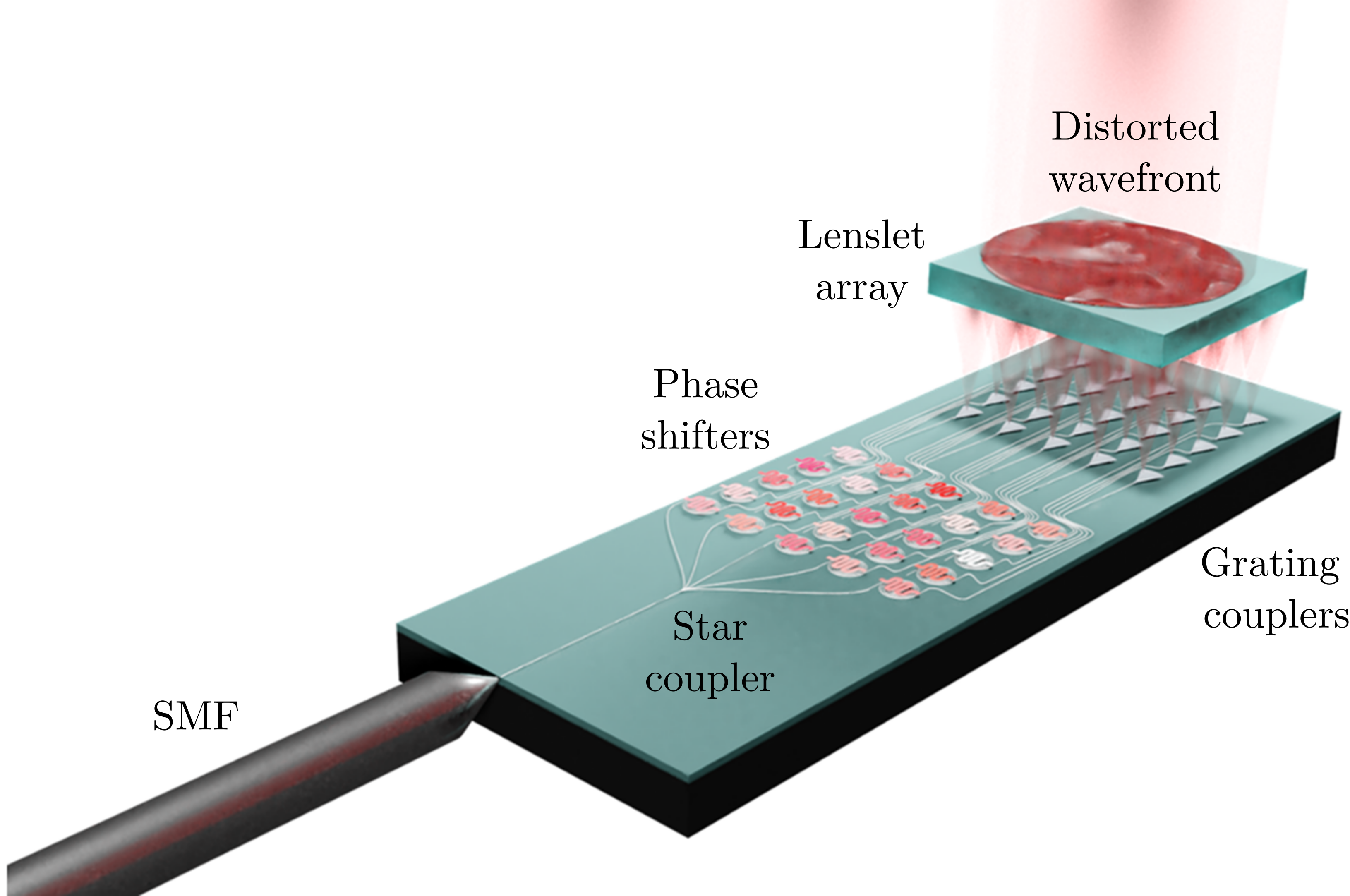}
    \caption{Concept of the photonic phase corrector. The distorted wavefront is focused by the lenslet array on the surface grating couplers. The coupled beamlets are shifted in phase and then combined before getting delivered to an SMF.}
    \label{fig:concept}
\end{figure}

As shown in Fig. \ref{fig:concept}, the PIC has a square array of grating couplers that inject light from free space into the plane of single-mode waveguides in a silicon-on-insulator (SOI) chip. High-resistance metal overlays heat long segments of the waveguides and modulate their refractive index through the thermo-optic effect. By shifting the phases of the propagating modes, the channels can be coherently combined, and the collected light can be delivered to one output waveguide. The co-phased beams are combined pairwise using Y-junctions in a binary tree architecture. A sub-wavelength grating (SWG) structure expands the mode near the end of the output waveguide, where the light can be edge-coupled out of the PIC to an SMF. 

\begin{figure}
    \centering
    \includegraphics[width=.8\linewidth]{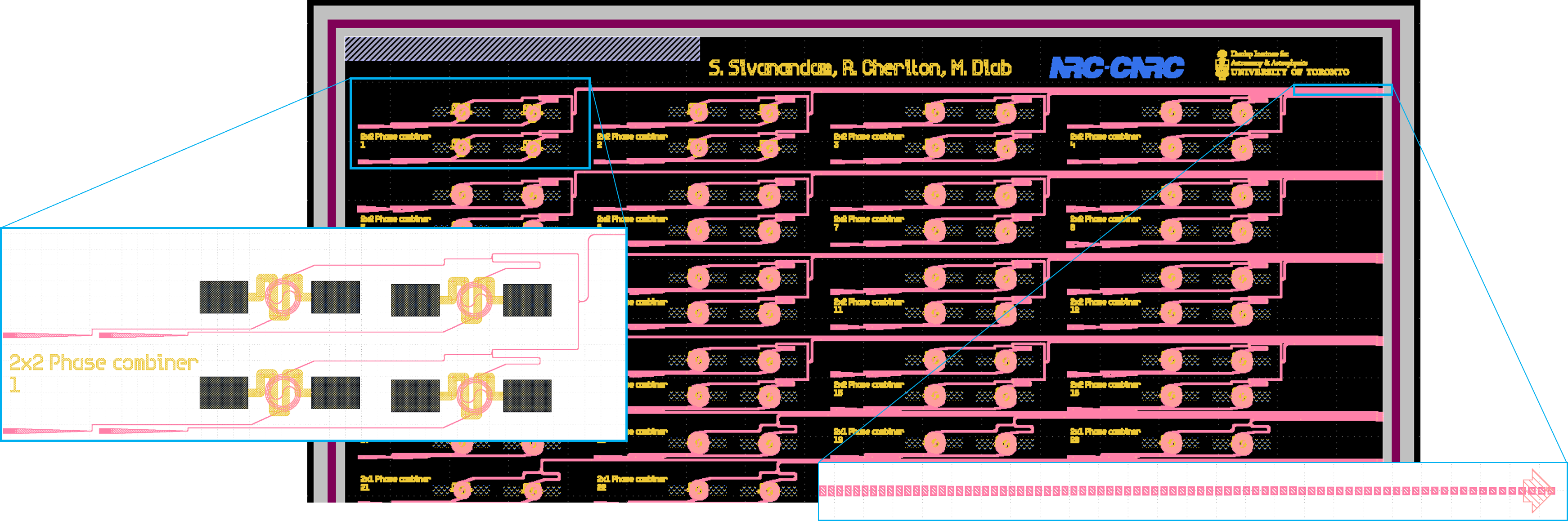}
    \caption{Part of the mask of the first-generation chip. Multiple PICs for different device sizes are included. The $32$ devices (only some are shown) are labeled in yellow. Also in yellow are the TiW heaters. Pink traces are the Si waveguides. The metal pads are depicted in gray. The insets show blowups of the first $2\times 2$ device and its SWG edge coupler.  }
    \label{fig:mask}
\end{figure}

The mask in Fig. \ref{fig:mask} shows the design of the $8.9$ mm $\times$ $8.9$ mm chip that contains several PICs.  
The design follows the standard process of e-beam lithography foundries. The chip was fabricated by the NanoSOI service from Applied Nanotools (ANT). The SOI chip has a $220$ nm thick silicon layer on a $2$ $\mu$m buried oxide (BOX) layer and a $725$ $\mu$m handle Si wafer. A $2.2$ $\mu$m oxide cladding (TOX) is deposited on top of the waveguides which also serves as protection from the environment. Titanium-tungsten (TiW) metal traces with a  $200$ nm thickness are deposited on top of the TOX layer to create the heaters, whereas titanium-tungsten/aluminum (TiW/Al) bi-layers are used to make the contact pads and the traces to the heaters. All heaters have a $\sim 60$ $\Omega$ resistance. A third oxide layer is deposited to prevent the oxidation of the heaters where windows are opened over the pads to allow wiring.

\begin{figure}
    \centering
    \includegraphics[width=.8\linewidth]{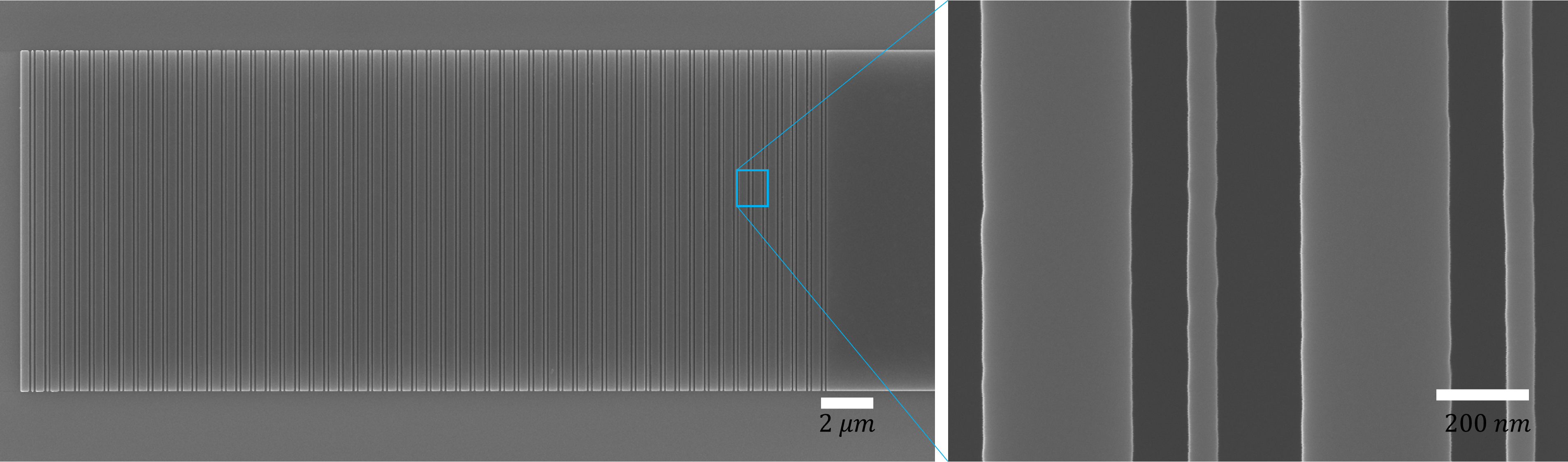}
    \caption{Scanning electron micrograph of the vertical incidence grating couplers. The blowup on the right shows the wall roughness of the grating lines.}
    \label{fig:grating_coupler}
\end{figure}

Different grating designs are used for the various devices in the PIC. The grating couplers that couple vertically-incident TE-polarized light at $1550$ nm are fabricated by a single-step etch process.  A high-resolution scanning electron micrograph of one of the gratings is shown in Fig. \ref{fig:grating_coupler}. The couplers have two interleaved periodic gratings, each with a pitch of 648 nm. An adiabatic tapered waveguide is downstream of each grating and matches the coupled beam to the mode of the $500$ nm single-mode waveguide. The phase shifters are coiled sections of the waveguides with the TiW metal layer deposited on top. The spiral geometry of the waveguides and the zigzag shape of the heaters maximize the available stroke. The stroke is estimated to be $>62$ $\mu$m at $20$ V. The subapertures are combined pairwise with $2\times 1$ combiners (Y-combiner) in a binary tree configuration.  

\begin{figure}
    \centering
    \includegraphics[width=.8\linewidth]{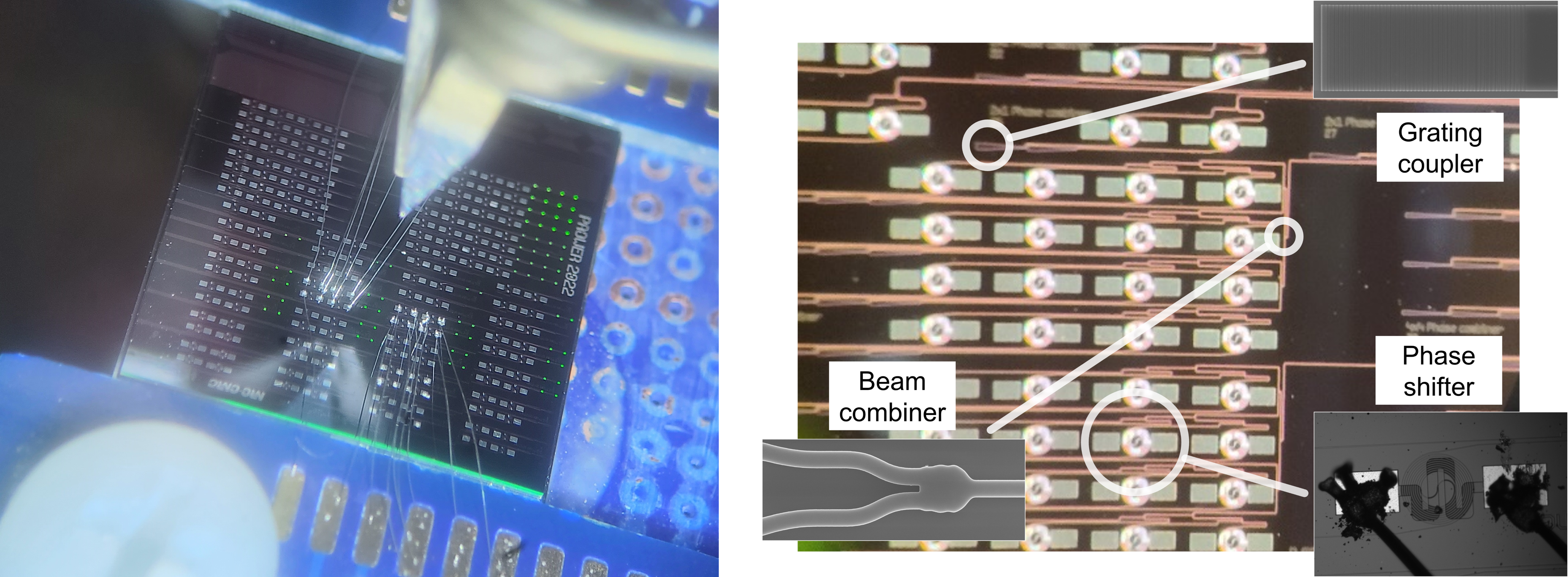}
    \caption{Left: Image of the wirebonded chip. The grating couplers can be identified by the diffracted green light from the white illumination source. Right: A magnified image of the circuit with micrographs of its components in the insets.}
    \label{fig:micrograph}
\end{figure}

The output waveguide from the tree is coupled to an SWG that expands the mode size to $3$ $\mu$m to allow edge coupling with a lensed SMF28 fiber \cite{cheben_swg_edge_coupler}. The mask includes a total of $32$ devices with different array sizes in addition to some propagation loss testing structures. Devices with $2\times 2$ subapertures are prioritized at this stage of the project, and $16$ of them are included. The three devices that have shown the best performance in terms of throughput, bandwidth, and central wavelength during the characterization were wire-bonded and used for further experiments. A printed circuit board (PCB) was designed to facilitate wire bonding and protect the chip facets (see Fig. \ref{fig:micrograph}).  

\section{Photonic phase correctors vs deformable mirrors}
\label{sec:dms}
Several technologies have been developed for DM implementation, but the most common are based on the use of the ferroelectric effect or voice coils to locally actuate and shape a thin mirror facesheet. Such DMs have high accuracy and speed, which is sufficient for most astronomical applications. However, they are costly and large in size, due to a large actuator pitch, which makes them unsuitable for multi-object applications. Alternatively, DMs based on MEMS technology are compact and consume less power, but they remain costly and have limited mechanical stroke, requiring an upstream correction of the larger low-order modes by another DM based on a different technology. Quantitatively, the residual wavefront error after correcting by a DM is estimated by \cite{madec_overview_2012}


\begin{equation}
    \label{eq:error}
    \sigma^2_{residual} = \sigma^2_{fitting} + \sigma^2_{temporal},
\end{equation}
where the fitting error results from the finite number of actuators and the temporal error results from their nonzero response time. Other error terms, such as the noise in the WFS signals and anisoplanatism, may be added, but the focus here is on the WFC properties. The fitting error decreases as the number of actuators $N$ increases, but it also depends on the arrangement of the actuators, their influence function, and the continuity of the mirror facesheet. The fitting error is given by \cite{noll_zernike_1976} 

\begin{equation}
\label{eq:fit}
    \sigma^2_{fitting} = \alpha \left( \frac{D}{r_0} \right)^{5/3} N^{-5/6},
\end{equation}
where $D$ is the telescope diameter and $r_0$ is the Fried parameter. The factor $\alpha$ depends on the geometry of the DM \cite{roddier_adaptive_1999}. The temporal error is given by

\begin{equation}
    \sigma^2_{temporal} =  \left( \frac{\tau \nu}{0.314 r_0} \right)^{5/3},
\end{equation}
where $\tau$ is the lag between when the command is applied, and the DM settling at the requested shape. The parameter $\nu$ is the magnitude of the effective wind velocity. The Strehl ratio ($\mathit{SR}$) of the corrected point spread function (PSF) at the focal plane drops exponentially as the residual error increases, i.e., $\mathit{SR} = \exp (-\sigma_{residual}^2) $.





To apply the same treatment as above to photonic WFCs, the concept of $\mathit{SR}$ must be expanded. The photonic WFC does not reduce the angular resolution of point sources in the image plane but rather improves the coupling efficiency into an SMF. However, since an SMF samples the center of the PSF and, in doing so, acts as an $\mathit{SR}$ sensor \cite{hubin_ground_2005, diab_starlight_2021}, the two functions are related and the same figure of merit can be used for both WFCs. The photonic Strehl ratio ($\mathit{SR}_{\mathit{ph}}$) is defined as the total power in the output SMF normalized by the total power in the fiber under diffraction-limited conditions. The $\mathit{SR}_{\mathit{ph}}$ of the photonic corrector can be established from finite-difference time-domain (FDTD) simulations. \cite{diab_photonic_2022, patel_end--end_2024} It is found that for the photonic WFC, the fitting error depends on $D/r_0$ and $N$ as the $5/6$-th power, instead of the $5/3$-rd as is the case for segmented DMs (see Eq. \ref{eq:fit}) 

\begin{equation}
\label{eq:fit2}
    \sigma^2_{ph, fitting} = \alpha \left( \frac{D}{r_0} \right)^{5/6} N^{-5/12}.
\end{equation}

The photonic WFC, however, is affected by an additional source of error that DMs are not. The non-uniformity in the field's amplitude at the pupil, resulting from free space propagation between the atmospheric layer and the telescope, i.e., scintillation, impacts the correction quality of the photonic WFC adversely. This adds another term, $\sigma^2_{scintillation}$, to the error budget of the photonic WFC in Eq. \ref{eq:error}. The temporal error is negligible in the case of a photonic WFC because of the fast-acting thermo-optic effect.

\section{Optical Setup}
\label{sec:setup}

\begin{figure}
    \centering
    \includegraphics[width=.6\linewidth]{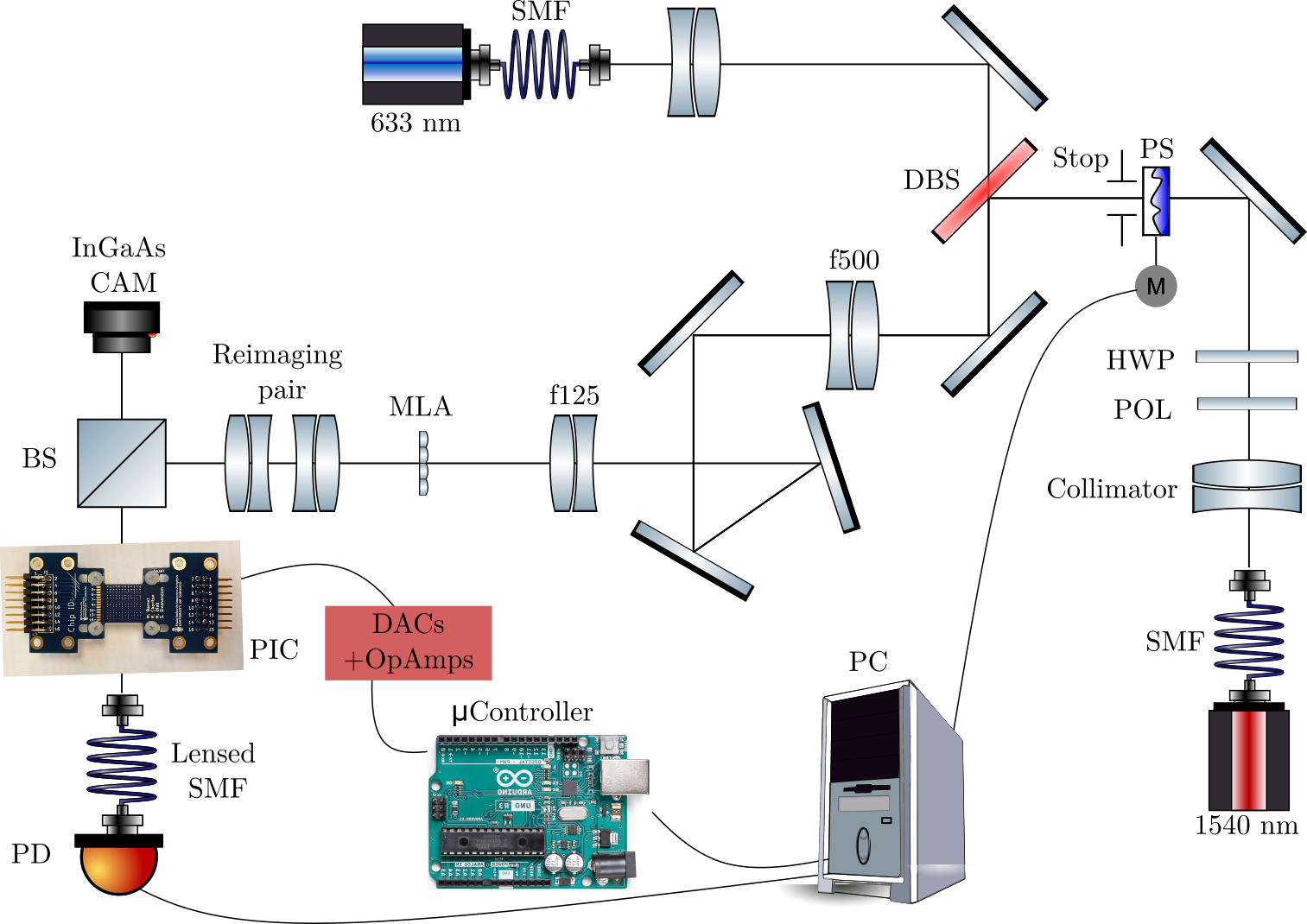}
    \caption{Schematic of the optical setup. A visible laser is used for alignment, while the $1540$ nm laser is the source coupled to the PIC. Polarization is controlled in free space, and a phase screen on a rotary stage introduces distortions in the beam that statistically match atmospheric turbulence. The beam is reduced to illuminate the required number of lenslets. A reimaging system separates the MLA and the PIC to allow viewing the spots, and the grating couplers for alignment.}
    \label{fig:schem}
\end{figure}

The optical setup to test the PIC uses a collimated laser beam and phase screens to emulate the received beam from a point source and the distorting atmospheric layers, respectively. As illustrated in Fig. \ref{fig:schem}, a polarized $1540$ nm fiber-coupled laser source is collimated with an achromat. A linear polarizer is used to analyze the polarization state, and a half-wave plate rotates the polarization to match the TE polarization of the grating couplers in the PIC. Phase screens with different turbulence strengths can be placed at the aperture stop of the system, where the beam is $3$ mm in diameter. A Keplerian telescope with a magnification factor of $4$ reduces the beam size to illuminate $2\times 2$ lenslets of the MLA at the pupil. An achromatic lens pair reimages the focal plane of the MLA on the grating couplers without magnification. This arrangement, together with a beam splitter, allows a downward-facing camera to view the spots on the surface of the PIC using a macro lens for alignment. The chip's PCB is mounted on a 6-axis precision micropositioner, used to align the PIC to the focal spots. The output fiber is also mounted on a micropositioner and can be aligned with the downward-facing camera. The off-the-shelf fiber is a single-mode lensed fiber with a $2.5$ $\mu$m mode field diameter (MFD), slightly smaller than the $3$ $\mu$m $1/e^{2}$ beam diameter at the chip facet. Table \ref{tab:setup} lists some of the parameters of the setup's components.

\begin{table}
    \centering
\caption{Optical setup components.}
\label{tab:setup}
    \begin{tabular}{|c|c|} \hline  
         LD power/wavelength& $40$ mW / $1540$ nm, polarized\\ \hline  
         MLA format/pitch/focal length& $30\times 30$ / $300$ $\mu$m / $1.7$ mm\\ \hline  
 Positioner resolution&X/Z$\sim10$ nm, Y$\sim 50$ nm, RYP$\sim 0.2$ arcsec\\ \hline  
 Output fiber&SMF28 (lensed)\\ \hline 
 PD sensitivity/resolution&$100$ pW / $10$ pW\\ \hline  
 DAC bandwidth/max. current&$70$ kHz / $60$ mA/channel\\ \hline 
 Microcontroller&Arduino Uno R3\\ \hline
    \end{tabular}
\end{table}

The thermo-optic phase shifters are driven using an Arduino microcontroller, digital-to-analog converters (DACs), and operational amplifiers (opamps). The circuit can drive the PIC at up to $70$ kHz while supplying up to $60$ mA of current to each heater. The power in the output fiber is measured by an InGaAs photodetector that feeds back the output signal to the PC which is used to close the loop and control the system.    

Three different phase screens are used to obtain lab data. With the beam diameter fixed at $3$ mm, the different $r_0$ values of the screens allow us to examine three different turbulence situations, namely $D/r_0 = 0.90, 1.59$, and $3.41$. The phase screens are mounted on rotary stages and can be rotated at variable speeds.
    
\begin{figure}
    \centering
    \includegraphics[width=\linewidth]{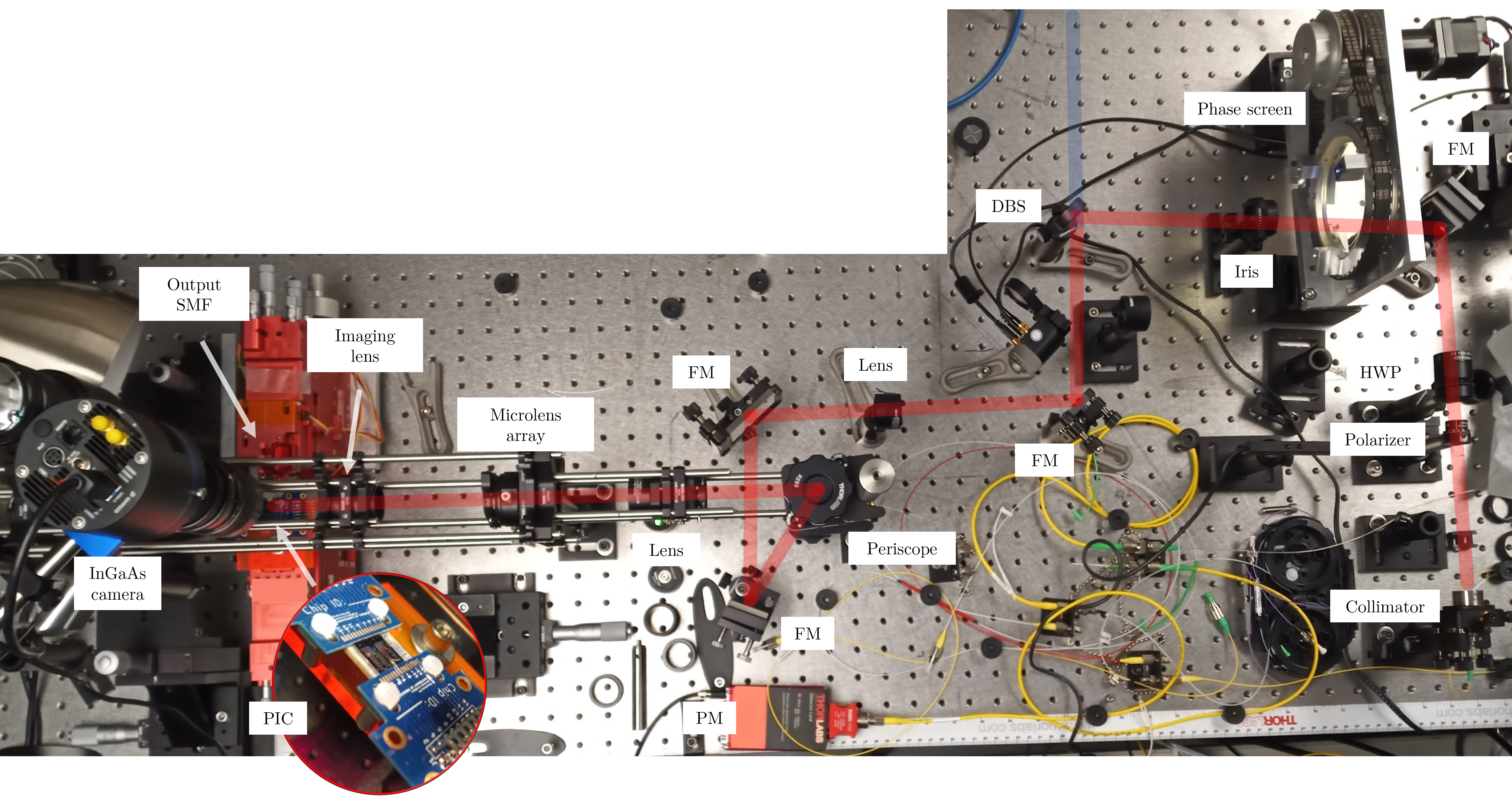}
    \caption{The optical setup for testing the PIC in the lab. The inset shows the PIC mounted on the micropositioner. FM: folding mirror. HWP: half-wave plate. PM: power meter. DBS: Dichroic beam splitter.
    }
    \label{fig:topview}
\end{figure}

A ray trace is used to calculate the pupil positions and assess the effect of optical aberrations on the focal spots. A top view of the setup in its final state is shown in Fig. \ref{fig:topview}.


\section{Experimental Results}
\label{sec:results}

\subsection{Phase shifters characteristics}
\begin{figure}
    \centering
    \includegraphics[width=0.5\linewidth]{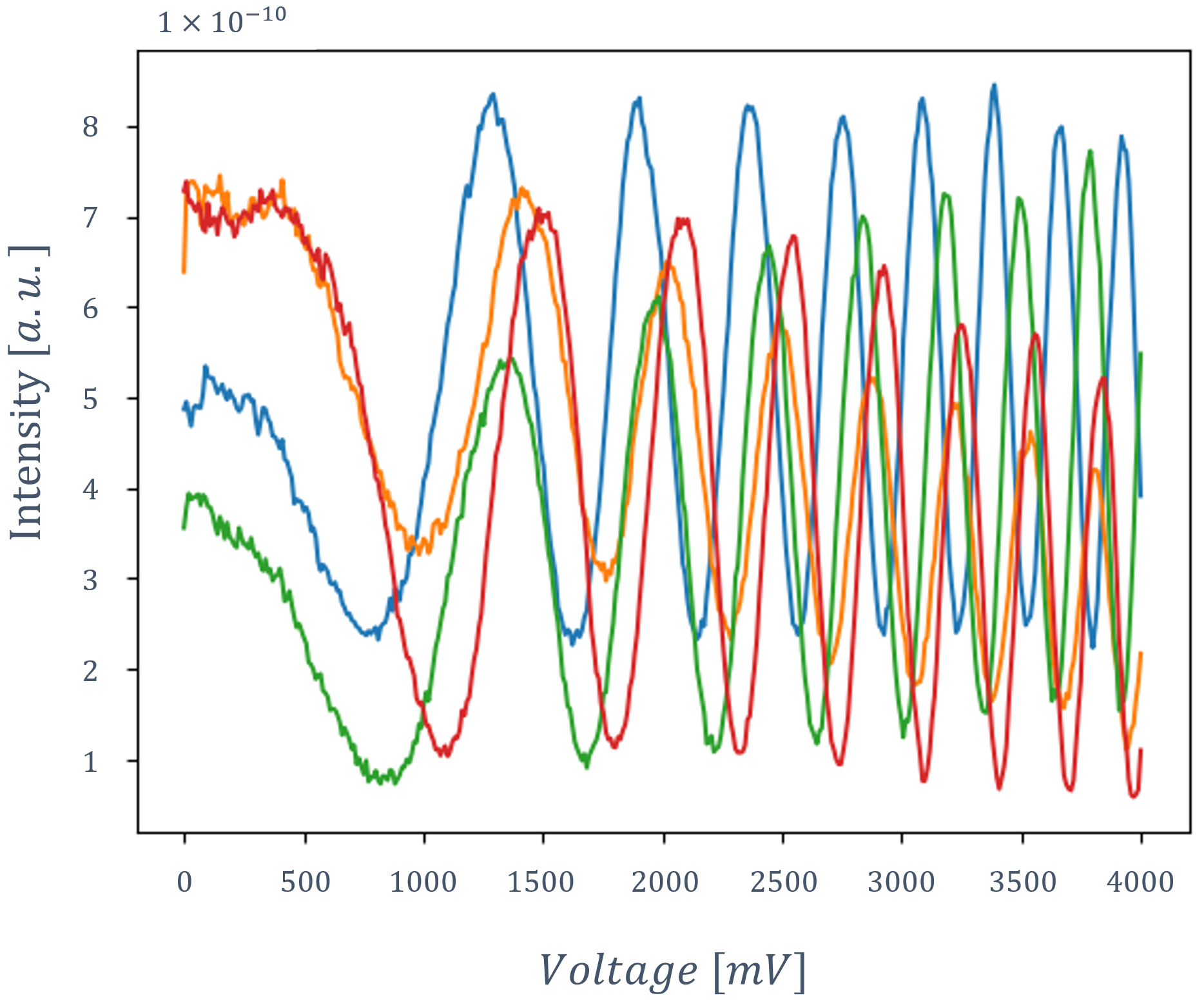}
    \caption{Ramp response graphs for a $2\times 2$ device. The graphs show the modulation in the output power as the voltage of one phase shifter is swept from $0$ to $4$ V while the other shifters are held at a constant voltage.}
    \label{fig:ramp}
\end{figure}

The first tests performed on the PIC were to understand its dynamic range and voltage response. Ramp response graphs (shown in Fig. \ref{fig:ramp}) record the optical power measured at the output SMF when the voltage across each heater is linearly increased from $0$ to $4$ V while holding the other inputs constant. The graphs reveal the "dead zone" at $V \lesssim 500$ mV where the supplied current, and hence the temperature change above room temperature, is not sufficient to cause a significant phase shift, i.e., change in the waveguide's refractive index. As the voltage increases, the refractive index modulation is observed as interference fringes in the output. The half-wave voltage, $V_{\pi}$, decreases with voltage as can be seen by the increase in the frequency of the fringes since heating becomes more efficient ($T \propto P \propto V^2$) at higher voltages. The nominal $V_{\pi}$ is about $250$ mV at our operating point ($V=2$ V, in Fig. \ref{fig:ramp}). The spiral segments of the phase shifters are $\sim 4$ mm long. Taking the thermo-optic coefficient to be $dn/dT = 2\times10^{-4}/$K, the temperature change caused by $V_{\pi}$ is about $1^{\circ}$C. The voltage could be increased to $20$ V to achieve longer strokes before any damage could be done to the PIC but the dependence of the resistance and the thermo-optic coefficient on temperature would cause a nonlinearity in the ramp response that requires careful calibration to operate over wide ranges. The variation in modulation depths between the different channels is attributed to optical path differences in free space between the beamlets. 

The magnitude of the stroke or optical path difference (OPD) needed to flatten a wavefront is proportional to $(D/r_0)^{5/6}$ and is achromatic since $r_0 \propto \lambda^{6/5}$. To compensate for $\pm 2.5 \sigma$, where $\sigma$ is the standard deviation of the wavefront OPD, the required stroke is \cite{hardy_adaptive_1998}  



\begin{equation}
    l = \frac{2.5 \lambda}{2 \pi} \left( \frac{D}{r_0} \right)^{5/6}.
\end{equation}

For applications where light is quasi-monochromatic, e.g., FSO communication links, phase wrapping means a phase shift of only $2\pi$ rad or one wave is sufficient. For an astronomical telescope with turbulence strength $D/r_0 = 8$, the required stroke is $2.3\lambda$ or $\sim 3.5$ $\mu$m at $\lambda = 1550$ nm. Therefore, the stroke supplied by the photonic WFC is sufficient to compensate for atmospheric turbulence in a mid-size telescope. However, the device is limited by the spectral bandwidth of the grating couplers as explained in Sec. \ref{sec:throughput}. 

\subsection{Throughput and spectral bandwidth}
\label{sec:throughput}

\begin{figure}[h]
    \centering
    \includegraphics[width=0.7\linewidth]{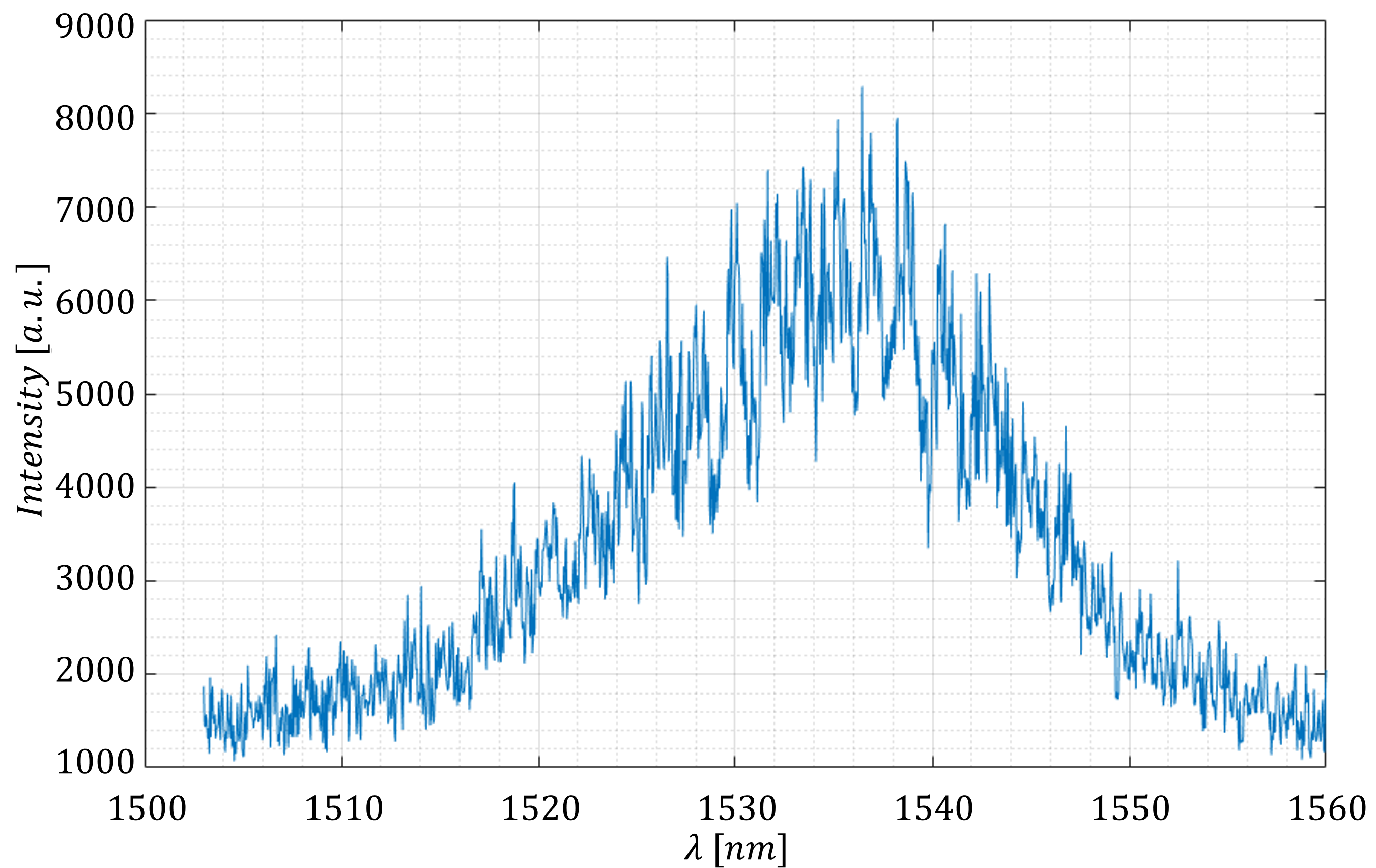}
    \caption{The spectral response measured for one device in the PIC with a C-band tunable laser. The slight skewness of the Gaussian spectrum to the right results from the uneven QE response of the detector used for the measurement. The fringes are due to the back-reflection from the Si handle.}
    \label{fig:spectrum}
\end{figure}

Figure \ref{fig:spectrum} shows the spectrum measured for a $2\times 2$ device. The curve shows an FWHM bandwidth of $\sim 25$ nm, significantly narrower than the $\sim 100$ nm band predicted by simulations \cite{diab_photonic_2022}. The central wavelength is $1535$ nm, which deviates from the $1550$ nm design wavelength and explains part of the loss in throughput we measured in the lab while using a $1540$ nm LD as a source. However, the performance of the system is significantly affected by the coupling of light in and out of the PIC. When coupling out, we used a lensed fiber that has a slightly narrower MFD than the optimum, and we aligned the fiber at a gap slightly larger than the working distance to avoid damaging the chip's facet. The losses from both of these imperfections in the setup were estimated from simulations. The propagation losses inside the PIC are difficult to isolate in the lab, and the estimates from the foundry are taken to be accurate. Most of the loss, however, takes place when coupling from free space into the PIC using the MLA and the grating couplers. The narrow bandwidth in Fig. \ref{fig:spectrum} hints at defects in the fabrication of the gratings, which we plan to investigate by running simulations with added tolerances to the grating geometry to identify the defect. Table \ref{tab:loss} gives the estimate for the potential throughput of the system with component and platform optimization.

\begin{table}[h]
    \centering
    \begin{tabularx}{\textwidth}{| p{0.3\textwidth} | p{0.1\textwidth} | X |}
    \hline
         Component &  Loss (dB) &     \\ \hline 
         Propagation loss& $-0.1$& Propagation Loss can be mitigated with smaller chips and better fabrication.  The propagation loss can be lower for ribbed silicon waveguides.\\ \hline 
         Phase delay lines & $-0.1$&  Losses in spiral can be mitigated by using shorter delay lengths with more efficient heaters. Improved fabrication can also improve the delay line loss.  \\ \hline 
         Grating couplers &  $-2$ &  Can be improved through a two-step etch process, back-reflector integration, and multi-layer waveguide integration, and optics for collimated and angled incident beams.  \\ \hline 
         Edge couplers &  $-0.2$&  Can be optimized to relax performance requirements for polarization independence and bandwidth.\\ \hline 
         Combiners &  $-1$&  Can be optimized using long structures and narrow bandwidths. \\ \hline 
         
 Projected device throughput after optimization& $-3.5$ (45\%)&\\ \hline
    \end{tabularx}
    \caption{Estimated projected throughput and optical losses with optimized components and optics.}
    \label{tab:loss}
\end{table}


\subsection{Phase correction and control}
From a control perspective, the photonic grating coupler array serves as a direct alternative to a conventional WFC. 
The chip can be controlled within a standard open-loop control setup, a control scheme analogous to several existing AO systems.
However, open-loop control requires a WFS, such as a SH-WFS. 
Although this component adds complexity and cost, it also imposes limitations on the control loop speed, as the WFS requires time for image formation.
Typically, this is not problematic for AO systems, as their DMs share similar response time constraints (around $1$ kHz). However, photonic wavefront correctors can operate much faster, at speeds up to $100$ kHz. 
In an open-loop controller, this substantial speed enhancement remains untapped, as the control system's speed is still constrained by the WFS exposure time. 
Consequently, we have chosen to explore wavefront sensorless control algorithms, which has significantly reduced the complexity of our performance testing.

\begin{figure}[t]
    \centering
    \includegraphics[width=0.7\linewidth]{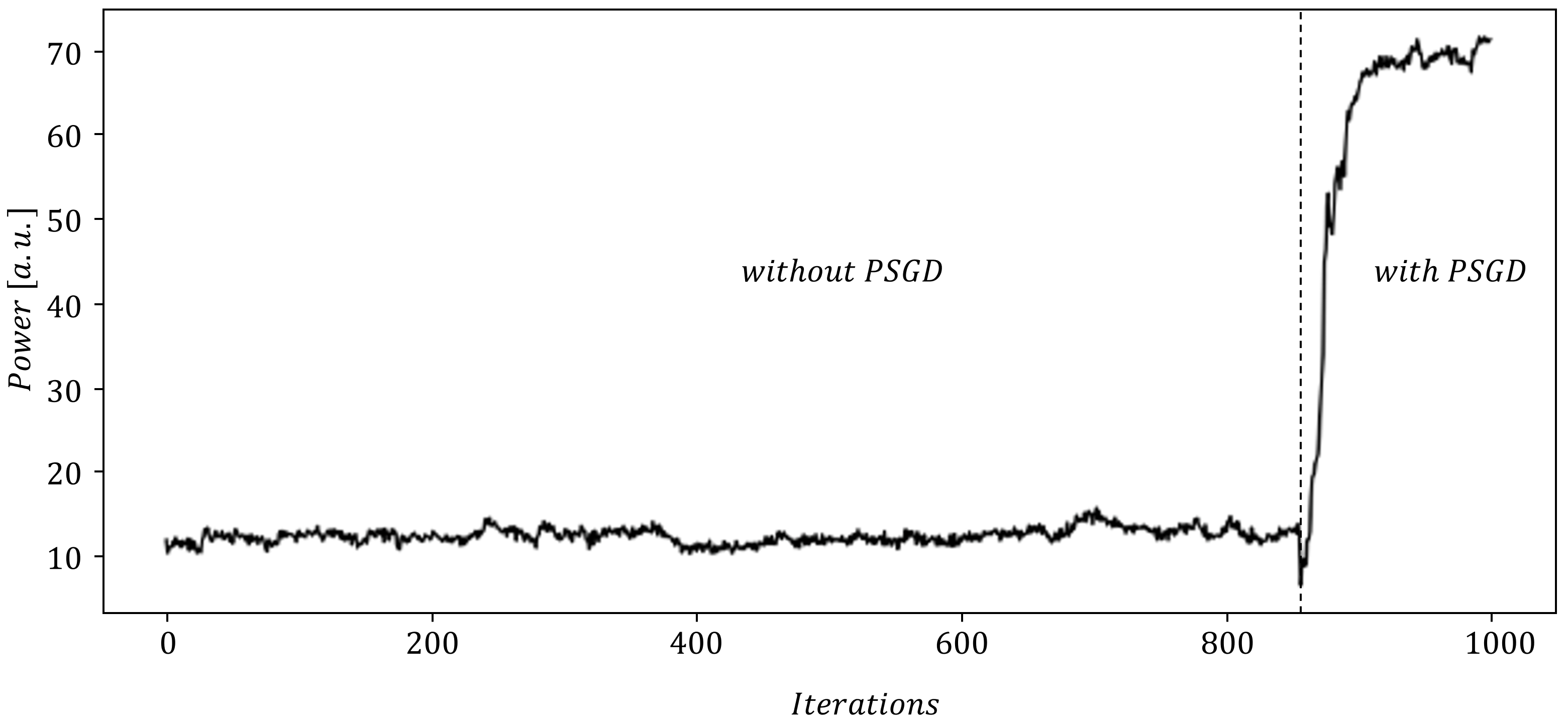}
    \caption{Closing the loop and phase locking on a phase screen with $D/r_0 = 1.59$.}
    \label{fig:close loop}
\end{figure}

\begin{figure}
    \centering
    \includegraphics[width=.7\linewidth]{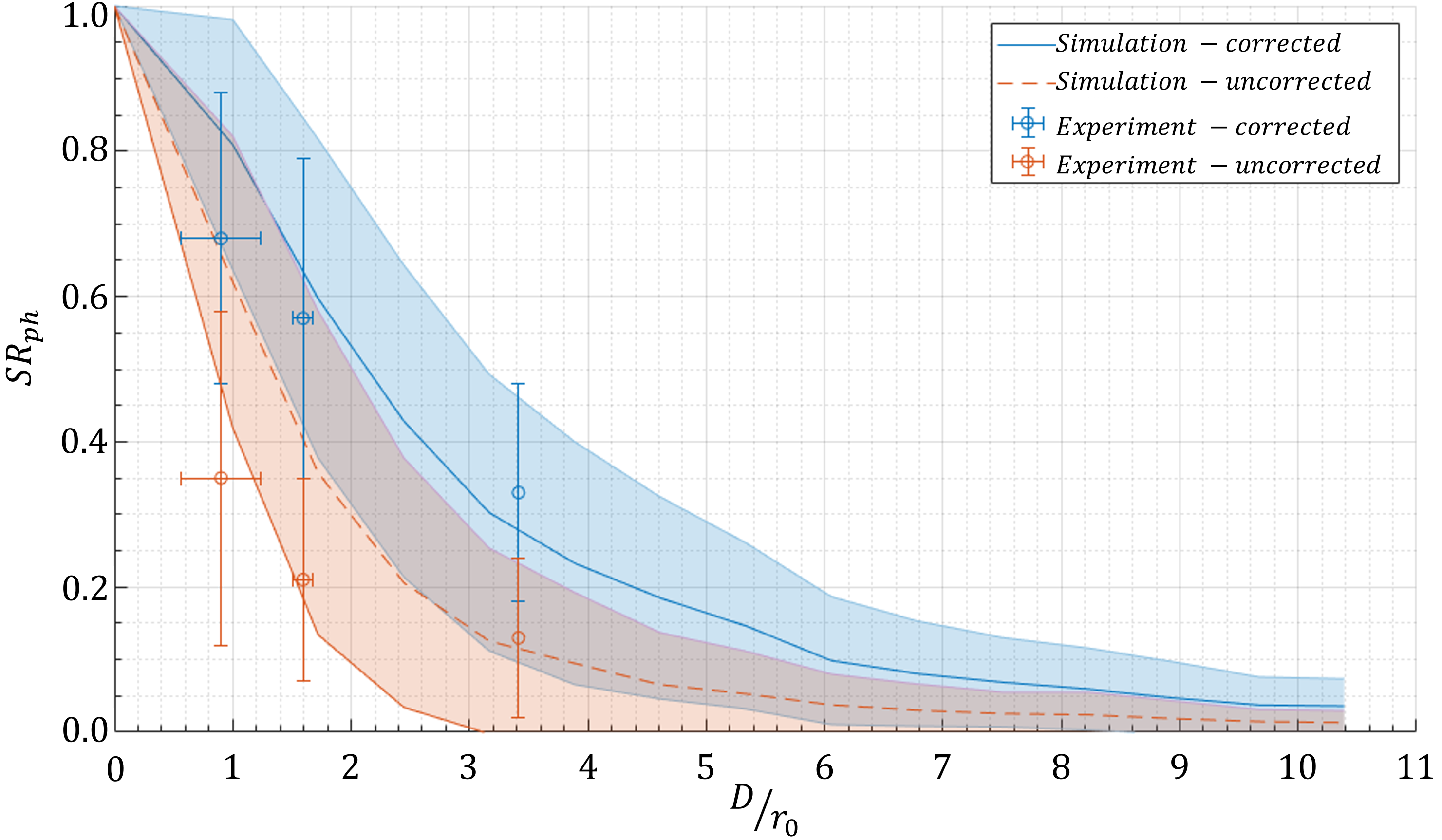}
    \caption{Plot of experimental data and the simulation prediction for $\mathit{SR}$ of a $2\times 2$ device.}
    \label{fig:Experiment_vs_Simulation}
\end{figure}

In this work, we have applied a sensorless wavefront controller to test our prototype $2\times 2$ grating coupler array.
Our wavefront sensorless controller implements the parallel stochastic gradient descent (PSGD)\cite{vorontsov_adaptive_2000} algorithm (Algorithm~\ref{alg:psgd}).
PSGD is particularly well suited to our system due to its scalability and small computational load, complimenting our significantly faster loop execution. 
Furthermore, PSGD can be implemented entirely within the chip's microcontroller, eliminating the need for external computer communication.
The chip's performance under wavefront sensorless control aligns closely with simulated performance under optimal control, providing a proof-of-concept for both the fundamental concept of the device and its control algorithm. Figure \ref{fig:close loop} shows the power in the fiber as the loop is closed, while
Fig.~\ref{fig:Experiment_vs_Simulation} compares the simulated performance of the PIC \cite{patel_end--end_2024} assuming an optimal controller with its empirical performance measured in the lab.


\begin{algorithm}[b]
\caption{Parallel Stochastic Gradient Descent}\label{alg:psgd}
\begin{algorithmic}[1]
\Procedure{PSGD}{$r$, $\epsilon$}
\State $i \gets 0$
\State $\alpha \gets 0$
\State $\setA \in \universe{N} \gets $ zero vector
\State $\setB \in \universe{N} \gets $ zero vector
\State $\mathbf{X}\in \universe{N} \gets $ zero vector
\While{True}
    \If{$i$ is Even}
        \State $\alpha \gets  \text{measureSignal()}$
        \State $\mathbf{X_n} \stackrel{iid}{\sim} U[-\epsilon, \epsilon]$ 
        \State $\setA \gets \setA + \mathbf{X}$
    \Else
        \State $\setB \gets (\text{measureSignal()} - \alpha) \times \mathbf{X}$ 
        \State $\setA \gets \setA + r\setB$
    \EndIf
    \State ApplyCommand($\setA$)
    \State $i \gets i+1$
\EndWhile
\EndProcedure
\end{algorithmic}
\end{algorithm}

Although potential improvements in wavefront sensorless control are evident, the feasibility of wavefront sensorless control under nominal on-sky conditions remains uncertain. 
PSGD fundamentally operates as a blind optimization algorithm, gradually optimizing a scalar output value with numerous control degrees of freedom. 
In our context, the number of grating couplers on the chip represents these control degrees of freedom, whereas photodiode measurements of light intensity serve as the signals we aim to optimize. 
For a given number of couplers and fixed atmospheric turbulence conditions, to fully correct the turbulence the number of iterations required to attain optimal signal strength, known as the rise time, must be shorter than the temporal coherence of our signal. 
As our output signal is only one dimensional, having more control degrees of freedom inherently demands more iterations to reach an optimum.
This requirement, coupled with the finite response time of the chip, raises practical concerns about the scalability of the algorithm to larger chip formats.

\section{Discussion and future work}
\label{sec:discussion}

The experimental results in Sec. \ref{sec:results} represent a proof-of-concept of the photonic WFC. The devices tested showed the ability to correct wavefront distortions to the extent predicted by end-to-end simulations\cite{patel_end--end_2024}. However, testing was limited to $2\times 2$ devices at this stage of the project. 
The experiments were also limited to turbulence strengths $D/r_0 < 3.4$ corresponding to the strongest phase screen available at our AO lab. Both limitations will be addressed in the future.  We did not measure the temporal bandwidth of the control loop by speeding up the rotary stage. The heaters have a fast response time, but the bandwidth is limited by the integration time of the photodetector, the speed of the serial communication used to command the DACs and read out the photodetector, and the convergence time of PSGD.

The SR is the performance metric of choice for astronomical AO systems, and we chose to use it here, too. It quantifies the efficiency of the PSGD algorithm and some aspects of the PIC, namely, the stroke and resolution of the phase shifters but does not account for the throughput of the optics. The throughput is quantified separately by measuring the ratio of power-in-fiber to the total power at the surface of the PIC. The two metrics, SR and throughput, can be multiplied to obtain an overall efficiency measure of how the photonic WFC enhances coupling in SMFs.

The agreement between the lab results and the simulation results validates our numerical models. The performance of the PIC is also within the expectations of SR estimated by the textbook formulas and rules of thumb used in astronomy. The empirical formulas estimate the residual wavefront errors and the SR given the strength of turbulence and the nature of the DM, assuming an ideal controller. Although our phase corrector is not a DM, we conjectured that its performance should be close to that of a square-segmented DM with piston-only control. The slight deviation from this assumption was established by the simulation results\cite{diab_photonic_2022, patel_end--end_2024}. 

Multiple improvements are planned for the next-generation PIC and optical system. We will arrange the grating couplers in a hexagonal array to increase the fill factor with the circular telescope pupil, and a hexagonal MLA will be used for coupling. We will increase the number of controlled subapertures to $8\times 8$, the number of degrees of freedom required to efficiently correct for turbulence within a mid-size telescope. 
The increase in subapertures will also require that a WFS be included to control the PIC. We plan to arrange all the electrical pads on the north/south sides of the next-generation chip while reserving the east side for the output waveguides that will be placed at the correct $127$ $\mu$m spacing, suitable for out-coupling to fiber arrays. Extra measures could also be implemented to increase the coupling efficiency of the grating couplers (up to $\sim 0.8$\cite{roelkens_high_2007}). These include apodization\cite{marchetti_high-efficiency_2017}, Si overlays\cite{vermeulen_high-efficiency_2010}, and Bragg reflector mirrors\cite{mekis_scaling_2012}. 

The parameter space of the $2\times 2$ devices is small enough for the sensorless optimization approach used with the current PAOWER PIC to co-phase the subapertures and demonstrate an increase in coupling for a slowly rotating phase plate. The next PIC will have devices with $8 \times 8$ subapertures and a WFS will be required to measure the relative phases and compute the heater commands. We started simulations for two wavefront sensing strategies. The first approach uses a SH-WFS similar to what is used in astronomical AO systems, while the second approach uses integrated Mach-Zehnder interferometers (MZIs) on the chip to do both the sensing and the combination simultaneously.    

\begin{figure}
    \centering
    \includegraphics[width=.6\linewidth]{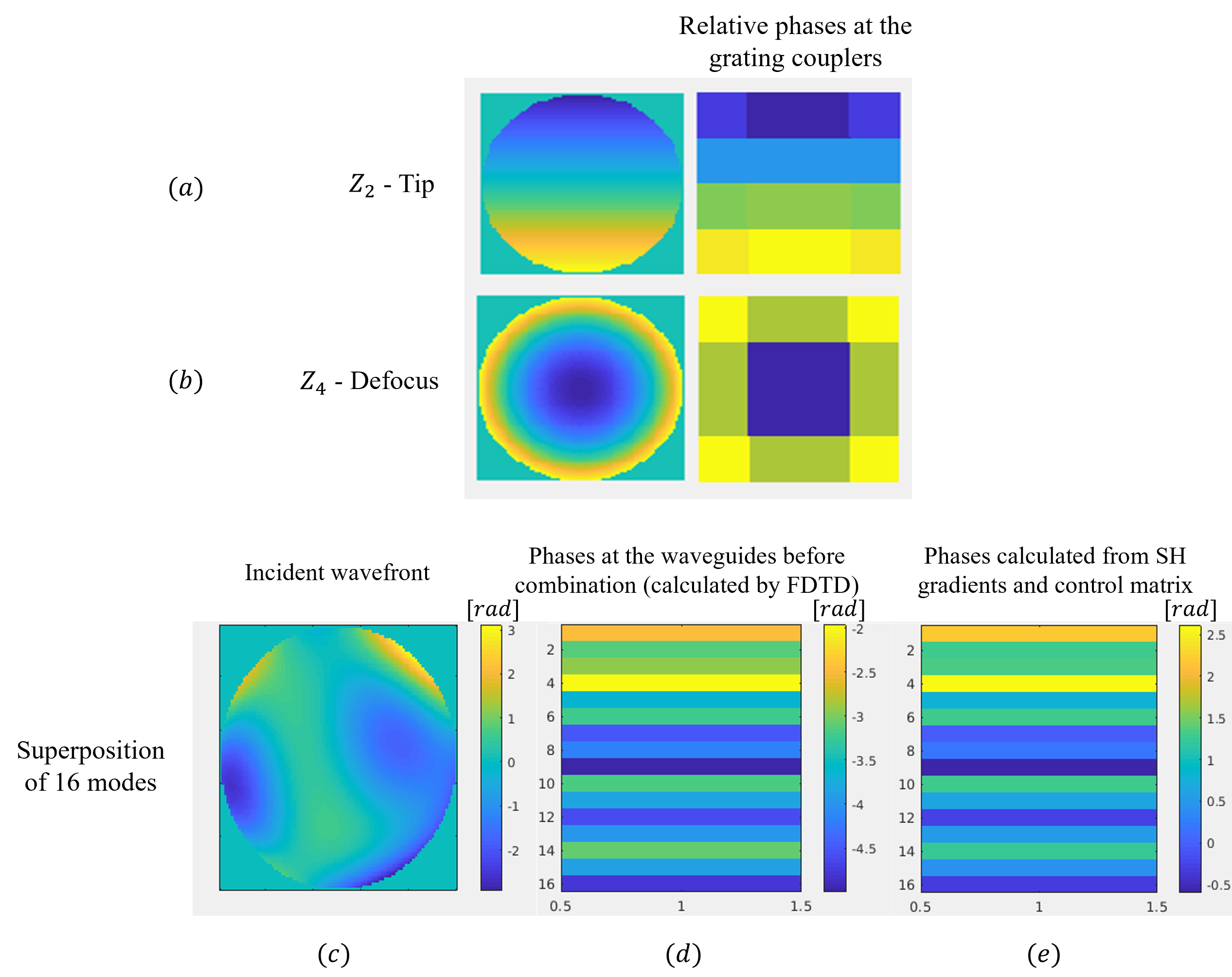}
    \caption{Example of a simulation result for control of a $4\times 4$ device with an SH-WFS. The relative phases of the beamlets coupled into the single-mode waveguides for two modes, (a) tip and (b) defocus. An example for the phase retrieval of an incident wavefront with $16$ modes (c). (d) The phases of the beamlets in the PIC calculated by an FDTD simulation. (e) The retrieved phases calculated from the SH-WFS and the control matrix.}
    \label{fig:SH-WFS}
\end{figure}

A SH-WFS used with the PAOWER PIC will have to measure the incoming wavefront before correcting it, unlike how it is normally done in astronomy where the WFS is in a closed loop and detects the residuals in the corrected wavefront downstream of the DM. In this latter configuration, the interaction matrix between the WFS measurements and the DM commands can be established by poking the DM actuators one by one or putting orthogonal shapes (usually Zernike modes) on the DM, while monitoring the WFS response to each command. For the photonic corrector, with the SH-WFS measurement prior to correction, the interaction matrix will have to be populated synthetically. A simulation of the propagation of the different Zernike modes through the chip will reveal the relative accumulated phases after coupling and before shifting. The SH-WFS measures the Zernike coefficients in the distorted wavefront, and the commands that compensate for that distortion can be computed using the two sets. Figure \ref{fig:SH-WFS} shows the relative phases at the shifters for two Zernike modes (tip and defocus) and an example of an incident wavefront showing the similar relative phases calculated from the FDTD simulations of the PIC and the propagation through the SH-WFS. The interaction matrix can also be established experimentally with the help of a DM or a spatial light modulator (SLM) in the lab. 

In the second approach, we will follow the combination concept reported by Miller \cite{miller_analyzing_2020} and Milanizadeh et al.\cite{milanizadeh_coherent_2021}. The beamlets from the subapertures are sent to a binary tree of MZIs that combines them two at a time by nulling the output in the sensing arm of each MZI. Simulations will need to be performed to understand the effects of scintillation and unequal intensities on the performance of MZIs as sensors/combiners. 

Packaging is also planned for the next-generation PIC. Packaging refers to the interfacing of the PIC with the coupling MLA, the output fiber, and the electrical connections. The first-generation chip was mounted on a thermally conductive segment of a PCB and wire-bonded to connection pads on each side. The MLA and output fiber were not packaged for this prototype, since we wanted to test multiple devices on the PIC, a few different fibers, and also study the system's sensitivity to alignment. Having a deconstructed setup also allows for measuring the throughput at different points in the optical train and inspecting the focal spots and beam profiles at the grating couplers and the chip's facet. In the future, we plan to align and mount the MLA directly on the surface and attach a fiber array at the output. Such packaging would produce a more robust system that requires minimum alignment and can be taken outside the lab for field tests.



\acknowledgments 
 
This work was supported by the High Throughput and Secure Networks Challenge Program of the National Research Council of Canada (HTSN 647 and 628). The authors also acknowledge CMC Microsystems for the provision of products and services that facilitated this research.

\bibliography{report} 
\bibliographystyle{spiebib} 

\end{document}